\documentclass[pre,twocolumn,showpacs,preprintnumbers,amsmath,amssymb]{revtex4}
\usepackage{graphicx}

\begin{document}

\title{Density-functional embedding using a plane-wave basis}

\author{J. R. Trail}\email{j.r.trail@bath.ac.uk}
\author{D. M. Bird}
\affiliation{Department of Physics, University of Bath, Bath BA2 7AY, UK}

\date{August, 2000}

\begin{abstract}
The constrained electron density method of embedding a Kohn-Sham system in a 
substrate system (first described by P. Cortona, Phys. Rev. B {\bf 44}, 8454 
(1991) and T.A. Wesolowski and A. Warshel, J. Phys. Chem {\bf 97}, 8050 
(1993)) is applied 
with a plane-wave basis and both local and non-local pseudopotentials.
This method divides the electron density of the system into substrate and 
embedded electron densities, the sum of which is the electron density of the 
system of interest.
Coupling between the substrate and embedded systems is achieved via approximate 
kinetic energy functionals.
Bulk aluminium is examined as a test case for which there is a strong 
interaction between the substrate and embedded systems.
A number of approximations to the kinetic-energy functional, both semi-local 
and non-local, are investigated.
It is found that Kohn-Sham results can be well reproduced using a non-local 
kinetic energy functional, with the total energy accurate to better than $0.1$ 
eV per atom and good agreement between the electron densities.
\end{abstract}

\pacs{71.15.Mb,71.15.Ap,71.15.-m}

\maketitle

\section{Introduction}
\label{sec:intro}
For the past two decades Density Functional Theory (DFT) 
\cite{jones89,dreizler90,parr89} has been one of the most powerful tools for 
the \emph{ab initio} calculation of the physical and chemical properties of 
materials.
Methods based on DFT make efficient use of computational resources, hence can 
generally deal with larger and  more complex systems than other \emph{ab 
initio} methods.
They also provide a simple interpretation of much of the many-electron physics 
of materials in terms of ideas based on the electron gas.
A number of implementations of DFT exist, which essentially differ in the 
approach taken to approximating the unknown density functional that describes 
the contribution to the electronic energy that is \emph{not} due to the 
external potential.

The most successful of these methods is the approach first derived by Kohn and 
Sham \cite{kohn65}, which uses DFT to identify the interacting electrons with a 
non-interacting electron gas, and then solves for the non-interacting electron 
system.
Most of the energy of the system is evaluated exactly, with only a relatively 
small exchange-correlation contribution requiring approximation.
In addition, the exchange-correlation part of the functional can be well 
approximated by a simple analytic form, hence this has become the workhorse of 
accurate DFT \cite{payne92}.
There is one main disadvantage that concerns us here.
Within the standard Kohn-Sham approach the electron density is expressed as the 
electron density of a non-interacting, many-electron system, by obtaining the 
eigenstates of these non-interacting electrons; this requires $O(N^3)$ 
operations where $N$ is the number of electrons present in the system.
It is this scaling behaviour that limits the size of system that the Kohn-Sham 
method can be applied to (currently less than around 1000 atoms).

A more direct and computationally cheaper approach is to minimise the total 
energy functional with respect to variations in the electron density, 
$\rho({\mathbf{r}})$ (eg Wang et al \cite{wang98}).
In this form the cost of finding the minimum of the total energy does not 
depend on $N$ provided the non-interacting kinetic energy functional 
$T_s[\rho]$ \cite{parr89} is available as an explicit functional of the 
electron density.
Unfortunately, this functional is not known, hence a direct minimisation 
procedure must employ approximate forms of the kinetic energy functional as 
well as the exchange-correlation energy.
Approximate expression are available, but as $T_s[\rho]$ is generally an order 
of magnitude greater than the exchange-correlation energy they are generally 
not of sufficient accuracy for structural optimisation, let alone chemical 
calculations.
Another deficiency is that there is no obvious way of applying the familiar 
non-local pseudopotentials of Kohn-Sham methods \cite{payne92} to these direct 
minimisation methods, although indirect methods have been proposed by Watson et 
al \cite{watson98}, Anata and Madden \cite{anata99} and Shah et al 
\cite{shah94}.

Other methods have been investigated, such as a path integral formulation of 
Kohn-Sham theory \cite{yang87}, and several approaches which formulate DFT as a 
$1^{st}$ order reduced density matrix theory \cite{dreizler90,goedecker98}.
The latter formulation takes advantage of the `nearsightedness' (see Kohn 
\cite{kohn96}) of the density matrix to solve for the ground state energy as an 
$O(N)$ problem, with the non-interacting kinetic energy functional evaluated 
exactly (eg Baroni and Giannozzi \cite{baroni92}, Hern\'{a}ndez et al 
\cite{hernandez96}, Ordejon \cite{ordejon98}, and a recent review by Goedecker 
\cite{goedecker99}).
Although \emph{ab initio} $O(N)$ approaches of this form are successful, they 
are currently limited to insulators as the density matrix is long-ranged for 
systems with no band gap.

A middle ground between the Kohn-Sham method and direct energy functional 
minimisation can be found via `embedding methods', and this is the approach 
considered in this paper.
In many cases the system we are interested in may be divided into two regions, 
$I$ and $II$.
Region $II$ is largely the same as a more simple system that may easily be 
solved for, whereas region $I$ is where the interesting physics occurs.
An example would be a defect in a crystal - region $II$ would be a bulk 
crystal, and region $I$ a small volume surrounding the defect.
There is an obvious computational advantage in solving for region $II$ first, 
and then solving for region $I$ taking into account the influence of region 
$II$ in some way.
This `embedding' approach has received a great deal of attention, and a large 
number of methods have been presented in the literature.
No attempt is given to review all of these, but we refer to publications that 
describe general classes of methods.
For the main part these differ in the space in which region $I$ and region $II$ 
are defined.
Inglesfield \cite{inglesfield81} defined these regions in real space and 
constructed an exact embedding scheme that requires a knowledge of the Green 
function of the substrate (region $II$), which is often prohibitively costly to 
calculate.
Regions $I$ and $II$ can also be defined more generally in Hilbert space, as 
described by Fisher \cite{fisher88}.
Another approach is discussed by Gutdeutsch et al 
\cite{gutdeutsch97,gutdeutsch98}, where the $1^{st}$ order reduced density 
matrix is partitioned.
All of these schemes are, to a greater or lesser extent, based on a 
wavefunction description of the embedding process.

Here we apply a different procedure, first described by Cortona 
\cite{cortona91} and Wesolowski and Warshel \cite{wesolowski93}, that 
carries out the embedding entirely at the density functional level.
The essential idea is to express the total energy of the system in terms of two 
Kohn-Sham like systems with densities $\rho_1({\mathbf{r}})$ and 
$\rho_2({\mathbf{r}})$ (corresponding to regions $I$ and $II$), where the total 
electron density is given by  
$\rho({\mathbf{r}})=\rho_1({\mathbf{r}})+\rho_2({\mathbf{r}})$.
This total energy is then minimised by varying only the electron density 
$\rho_1({\mathbf{r}})$, corresponding to fewer electrons than the whole system, 
and so a cheaper calculation.
Potential applications of this method include defects in crystals, or 
adsorbates on surfaces, since once a solution for the substrate is available 
the remainder of the calculation would only involve the Kohn-Sham 
representation of the electrons in the immediate vicinity of the defect or 
adsorbate.
This partially frozen electron density method has been implemented by Cortona 
\cite{cortona91}, Wesolowski et al 
\cite{wesolowski93,wesolowski94,wesolowski96,wesolowski99} 
and (in a slightly different context) 
Govind et al\cite{govind99} in a form that describes the coupling between 
the two subsystems via approximate kinetic energy functionals.
A localised set of basis functions was used in each case.

In this paper this procedure is implemented within a plane-wave pseudopotential 
framework \cite{payne92}.
We investigate the approach for metallic systems where the electrons in regions 
$I$ and $II$ are strongly interacting, whereas past applications have focused 
on insulating systems and a relatively weak interaction.
Our goal is to develop an embedding approach accurate enough, and 
computationally cheap enough, to aid the investigation of large scale defect 
and adsorbate systems.
As a preliminary to this we test and analyse the approach by constructing a 
four atom cell of bulk fcc aluminium by embedding one cubic sub-lattice of 
atoms within three others.
Sections \ref{sec:embthe} and \ref{sec:appke} describe the implementation of 
the method, with approximate kinetic energy functionals described from the 
viewpoint of a plane-wave basis.
Subsection \ref{sec:approxrho1} presents the results of the method applied to  
bulk fcc aluminium, with  a number of different kinetic energy functionals.
In subsection \ref{sec:exactrho1} a modified form of the method is applied in 
order to analyse the source of errors, again for different functionals.
Another important issue in the direct application of DFT methods is the 
inclusion of non-local pseudopotentials.
In subsection \ref{sec:nonloc} we present and justify the assumptions that 
must be made in order to employ non-local pseudopotentials within the 
embedding scheme.
Rydberg atomic units are used throughout unless otherwise stated.

\section{Partially Frozen electron density}
\label{sec:embthe}
We begin with the familiar Hohenberg-Kohn total energy functional $E[\rho]$, 
expressed in the Kohn-Sham form \cite{parr89}
\begin{equation}
E[\rho]=T_s[\rho]+
J[\rho]+E_{xc}[\rho]+\int V_{ext}({\mathbf{r}}) \rho({\mathbf{r}}) 
d^3{\mathbf{r}}
\label{e2.1}
\end{equation}
where $T_s[\rho]$ is the non-interacting kinetic energy functional, $J[\rho]$ 
is the Hartree energy, $V_{ext}$ is the local external potential (for 
the systems considered here it is given by the sum of local 
pseudopotentials at each atomic site) and $E_{xc}[\rho]$ is the 
exchange-correlation energy, including the kinetic-correlation contribution.
In the standard Kohn-Sham methodology the functional derivative of Eq. 
(\ref{e2.1}) is taken with the total number of electrons constrained to be 
constant.
By setting this equal to zero an Euler-Lagrange equation is obtained, which is 
identified with the Euler-Lagrange equation for a system of non-interacting 
electrons in a specific external potential.
Solving for this `reference' system to yield the same electron density as the 
interacting system is the essence of the Kohn-Sham implementation of density 
functional theory, and results in the electron density that gives the correct 
minimum of Eq. (\ref{e2.1}), the ground state energy.

The partially frozen electron density method breaks down this same functional 
into \emph{two} non-interacting gases.
The electron density of the system, $\rho({\mathbf{r}})$, is divided into two 
components so that 
$\rho({\mathbf{r}})=\rho_1({\mathbf{r}})+\rho_2({\mathbf{r}})$.
One of these components (in what follows, $\rho_2({\mathbf{r}})$) is taken to 
represent a part of the system that is expected to change very little (this 
will be qualified further on); the \emph{substrate}.
This is kept constant, and in what follows is obtained from a Kohn-Sham 
calculation so the kinetic energy $T_s[\rho_2]$ is known accurately.
The total energy functional is then expressed in terms of the kinetic energy 
functional of $\rho_1({\mathbf{r}})$ and an `embedding kinetic energy' term, 
$T^{nadd}_s[\rho_1,\rho_2] + T_s[\rho_2]$ that takes into account the influence 
of the rest of the system.
This gives the total energy as \cite{cortona91,wesolowski93}
\begin{eqnarray}
E[\rho]=&&T_s[\rho_1]+T^{nadd}_s[\rho_1,\rho_2]+T_s[\rho_2] \nonumber\\
&&\mbox{} +J[\rho]+E_{xc}[\rho]+\int V_{ext}({\mathbf{r}}) \rho({\mathbf{r}}) 
d^3{\mathbf{r}}
\label{e2.2}
\end{eqnarray}
where the non-additive part of $T_s[\rho]$, $T^{nadd}_s[\rho_1,\rho_2]$, is 
defined as
\begin{equation}
T^{nadd}_s[\rho_1,\rho_2]=
T_s[\rho_1+\rho_2]-T_s[\rho_1]-T_s[\rho_2].
\label{e2.3}
\end{equation}

Minimising Eq. (\ref{e2.2}) with respect to variations in 
$\rho_1({\mathbf{r}})$ only, with the substrate density $\rho_2({\mathbf{r}})$ 
constant, and a constraint of constant total number of electrons in 
$\rho_1({\mathbf r})$, results in the Euler-Lagrange equation
\begin{equation}
\frac{\delta T_s[\rho_1]}{\delta \rho_1} +
\frac{\delta T^{nadd}_s[\rho_1,\rho_2]}{\delta \rho_1}
+V_{KS}[\rho;{\mathbf{r}}]=\mu
\label{e2.4}
\end{equation}
where $\mu$ is an arbitrary constant reflecting the fact that for a fixed 
number of electrons the functional derivative is defined to within an additive 
constant only.
In the same manner as for the Kohn-Sham case, this leads to the 
$\rho_1({\mathbf{r}})$ being the solution of the `Kohn-Sham' equations 
associated with Eq. (\ref{e2.4}) at self consistency, but with an effective 
potential given by
\begin{eqnarray}
V^{eff}[\rho;{\mathbf{r}}]&&=V_{KS}[\rho;{\mathbf{r}}] + 
\frac{\delta T^{nadd}_s[\rho_1,\rho_2]}{\delta \rho_1} \nonumber\\
&&=V_{KS}[\rho;{\mathbf{r}}] +
\left(
\frac{\delta T_s[\rho_1+\rho_2]}{\delta \rho_1} - \frac{\delta 
T_s[\rho_1]}{\delta \rho_1}
\right).
\label{e2.5}
\end{eqnarray}
If an exact expression for the kinetic energy functional was available this 
prescription would provide a ground state energy and electron density exactly 
equivalent to the Kohn-Sham scheme for the total system, with one 
additional limitation.
Since $\rho_1({\mathbf{r}})$ takes the form
\begin{equation}
\rho_1({\mathbf{r}})=\sum_i w_i |\psi_i ({\mathbf{r}})|^2
\label{e2.6}
\end{equation}
it is positive, hence the trial densities that are searched to minimise 
the total energy in Eq. (\ref{e2.1}) satisfy 
$\rho({\mathbf{r}}) \geq \rho_2({\mathbf{r}})$, and the true ground state 
energy and density is obtained only if the ground state density satisfies this 
inequality.
In practise $\rho_2({\mathbf{r}})$ is chosen such that this is not a 
significant restriction, and this is true for the test cases considered in 
section \ref{sec:results}.
This constraint may also be relaxed by applying a `Freeze and Thaw' procedure, 
as discussed at the end of subsection \ref{sec:exactrho1}.

Since $\rho_2({\mathbf{r}})$ is taken as already known, this method of 
obtaining the electronic structure need only solve for the electrons present in 
$\rho_1({\mathbf{r}})$, a smaller number (in many cases considerably smaller) 
than is present in the entire system.
However, in order to apply this approach the term $T^{nadd}_s[\rho_1,\rho_2]$, 
or \emph{non-additive kinetic energy}\cite{lacks94,perdew88}, and its 
functional derivative in Eq. (\ref{e2.2}) and Eq. (\ref{e2.5}) are required.
Since no explicit form is available approximate kinetic energy functionals are 
employed in Eq. (\ref{e2.3}) to provide an approximate non-additive kinetic 
energy functional.
To clarify, we approximate all of the functionals on the RHS of Eq. 
(\ref{e2.3}), whereas in Eq. (\ref{e2.2}) the functionals $T_s[\rho_1]$ and 
$T_s[\rho_2]$ are exact.
This approximation to the non-additive kinetic energy is the only additional 
source of error introduced by the method, but $T^{nadd}_s$ is expected to be 
far smaller than the total kinetic energy (it is zero if $\rho_1({\mathbf{r}})$ 
and $\rho_2({\mathbf{r}})$ do not overlap) for most reasonable divisions of the 
electron density, and it could be hoped that some error cancellation will occur.

The non-additive kinetic energy has been investigated as a test of the quality 
of a number of kinetic energy functionals by Lacks and Gordon \cite{lacks94}, 
who compared $T^{nadd}_s$ for Helium and Neon calculated from approximate 
functionals with Hartree-Fock results.
They conclude that the fractional error of $T^{nadd}_s$ is greater than for the 
kinetic energy of the whole system, implying that for weakly interacting 
subsystems the error cancellation is limited.
Whether this is the case for more strongly interacting subsystems and for the 
functionals applied here will have a direct influence on the accuracy of our 
results.
In addition it should be remembered that even if the approximate functional 
gives the correct kinetic energy for the true density, this may not be a 
minimum of the total energy with respect to variations in the density. 

\section{Approximate Kinetic Energy Functionals}
\label{sec:appke}
Approximations to the kinetic energy functionals are used to construct 
$T^{nadd}[\rho_1,\rho_2]$ in Eqs. (\ref{e2.2}$-$\ref{e2.5}).
Many are available in the literature (eg Thakkar \cite{thakkar92}, Wang et al 
\cite{wang98}, Garc\'{i}a-Gonz\'{a}lez et al \cite{garcia96}, Herring 
\cite{herring86}), and these have been assessed in a number of environments 
ranging from isolated atoms to bulk systems to molecular interactions and 
surfaces.
The majority of these assessments investigate the ability of the functionals to 
reproduce accurate kinetic energies from accurate electron densities calculated 
by other means.
A smaller number of studies have examined the ability of approximate 
functionals to produce accurate electron densities and energies when the total 
energy is minimised using the functionals themselves, and little work has been 
published on the success of these approximations in reproducing accurate 
functional derivatives.
We have therefore chosen to examine a range of functionals.

Although an analytic gradient expansion exists for $T_s[\rho]$, convergence 
cannot be achieved for systems where the density decays exponentially 
\cite{dreizler90}.
This has been attributed to the expansion taking the form of an asymptotic 
series, as described by Pearson and Gordon \cite{pearson85}, although this has 
not been shown analytically.
In order to overcome this difficulty in improving the local density 
approximation (LDA) to $T_s[\rho]$, many authors have taken a similar approach 
to the Generalised Gradient Approximation (GGA) to the exchange-correlation 
functional, by carrying out a partial re-summation via an enhancement factor  
\cite{thakkar92}.
The kinetic energy is approximated by
\begin{equation}
T^{app}_s[\rho]=\frac{3}{5}(3\pi^2)^\frac{2}{3} \int
\rho^\frac{5}{3} F(t) d^3{\mathbf{r}}
\label{e2.7}
\end{equation}
where
\begin{equation}
t=\frac{|\nabla \rho|^2}{\rho^\frac{8}{3}},
\label{e2.8}
\end{equation}
hence the kinetic energy is expressed as a semi-local functional of the 
electron density and its gradient.
The functional derivative of a general semi-local density functional
\begin{equation}
G[\rho]=\int g(\rho,\nabla \rho, \nabla^2 \rho, \cdots) d^3{\mathbf{r}}
\label{e2.9}
\end{equation}
is given by \cite{parr89}
\begin{equation}
\frac{\delta G}{\delta \rho}=
           \frac{\partial g}{\partial          \rho} - 
\nabla  .  \frac{\partial g}{\partial \nabla   \rho} +
\nabla^2   \frac{\partial g}{\partial \nabla^2 \rho} - \cdots
\label{e2.10}
\end{equation}
which may be truncated at second order for the functional given in Eq. 
(\ref{e2.7}), since the kernel is a function of $\rho$ and $\nabla \rho$ only.

The direct application of Eq. (\ref{e2.10}) to the semi-local functional, Eq. 
(\ref{e2.7}), yields
\begin{eqnarray}
\frac{\delta T^{app}_s[\rho]}{\delta \rho}=\frac{1}{5}(3\pi^2)^\frac{2}{3}
&&\left[ 
5 \rho^\frac{5}{3} F(t) - 6 \frac{ \nabla^2 \rho}{\rho} F'(t) \right. 
\nonumber\\
&&\left. \mbox{} - 2 \frac{|\nabla \rho|^2}{\rho^2}(F'(t)-8tF''(t))  \right. 
\nonumber\\
&&\left. \mbox{} - 12 t F''(t) \frac{\nabla \rho . \nabla |\nabla \rho|}{\rho 
|\nabla \rho|}
\right]
\label{e2.11}
\end{eqnarray}
an unwieldy expression that involves highly non-linear terms.
With a plane-wave basis difficulties arise due to $\rho({\mathbf r})$ being 
defined on a real space grid, with gradients conventionally obtained via the 
Fast Fourier Transform (FFT).
Since Eq. (\ref{e2.11}) is non-polynomial in $\rho({\mathbf r})$ it cannot be 
represented by a finite Fourier space, and using any finite space results in 
aliasing errors.
For example, the term $\nabla \rho .\nabla|\nabla \rho|$ results in 
particularly large errors, and if Eq. (\ref{e2.11}) is applied directly this 
causes errors in the resulting potential to propagate through further 
iterations, preventing convergence.

Exactly this problem manifests itself in the application of the GGA with a 
plane-wave basis, as discussed by White and Bird\cite{white94}.
In this case convergence is also affected, though not as severely due to the 
exchange-correlation energy being an order of magnitude smaller than the 
kinetic energy.
We follow the same approach as White and Bird to find a more stable expression 
for the kinetic energy functional derivative.
The functional is discretised as a sum of contributions at each real space grid 
point,
\begin{equation}
T^{app}[\rho]=\frac{3}{5}(3\pi^2)^{\frac{2}{3}} \frac{\Omega}{N} \sum_{\mathbf 
R} \rho^{\frac{5}{3}}_{\mathbf R} F(t_{\mathbf R}),
\label{e2.12}
\end{equation}
where $\Omega$ is the volume of the unit cell, $N$ is the number of grid 
points, and the subscript $\mathbf R$ denotes the quantity at the grid point 
$\mathbf R$.
The `functional derivative' (in fact a total derivative) can then be written in 
terms of the partial derivative of Eq. (\ref{e2.12}) with respect to 
$\rho_{\mathbf R}$ and $\nabla \rho_{\mathbf R}$, which are considered as 
independent variables.
This yields the expression
\begin{eqnarray}
\left. \frac{\delta T^{app}_s[\rho]}{\delta \rho} \right|_{\mathbf R}  =
\frac{3}{5}(3\pi^2)^\frac{2}{3}
&&\left[
 \frac{5}{3} \rho_{\mathbf R}^\frac{2}{3} F(t)
-\frac{8}{3} \frac{|\nabla \rho_{\mathbf R}|^2}{\rho^2} F'(t) \right. 
\nonumber\\
&&\left. \mbox{} -\nabla . \left( 2 \frac{\nabla \rho_{\mathbf 
R}}{\rho_{\mathbf R}} F'(t) \right)
\right]
\label{e2.13}
\end{eqnarray}
which is analytically equivalent to Eq. (\ref{e2.11}) in the limit $N 
\rightarrow \infty$.
Expression (\ref{e2.13}) for the functional derivative is applied in our 
calculations, with all gradients calculated via the FFT, and is found to remove 
the instability inherent in Eq. (\ref{e2.11}). 

It is important to note that although Eq. (\ref{e2.13}) defines the functional 
derivative in a manner consistent with a finite real space grid, it does not 
ensure that an accurate representation of the functional derivative is obtained 
for a given electron density.
This is immediately apparent if an exponentially decaying electron density is 
considered.
In this case the second and third terms in Eq. (\ref{e2.13}) should result in 
the expressions $\frac{|\nabla \rho_{\mathbf R}|^2}{\rho^2_{\mathbf R}}$ and 
$\frac{\nabla \rho_{\mathbf R}}{\rho_{\mathbf R}}$ being constant due to a 
cancellation of the exponentials.
However, since the numerator in both these terms will in fact be the gradient 
of a trigonometric representation of the electron density an exact cancellation 
will not occur, and for small values of $\rho_{\mathbf R}$ errors in these 
terms (and the resulting functional derivative) will not be small.
This is found to prevent convergence if the derivative of $F(t)$ in Eq. 
(\ref{e2.13}) is large, but has no effect for the functionals considered here.

A number of enhancement factors available in the literature 
\cite{wesolowski96,thakkar92} were investigated (specifically the Thomas-Fermi 
approximation and $1^{st}$ order gradient expansion \cite{dreizler90}, and the 
functionals constructed by Thakkar \cite{thakkar92}, Vitos et al 
\cite{vitos98}, DePristo and Kress \cite{depristo87}, Ou-Yang and Levy 
\cite{ouyang91}, Lee et al \cite{lee91}, Perdew and Wang 
\cite{perdew86a,perdew86b,perdew91} and Lembarki and Chermette 
\cite{lembarki94}), but their overall behaviour was found to be similar.
We therefore present results for two enhancement factors only.
The first is that of Perdew and Wang, $F_{PW86}$ \cite{perdew86b},
\begin{equation}
F_{PW86}(s)=(1+1.296s^2+14s^4+0.2s^6)^{\frac{1}{15}},
\label{e2.14}
\end{equation}
where
\begin{equation}
s=\frac{1}{2 (3 \pi^2)^{ \frac{1}{3} } } t^{ \frac{1}{2}}.
\label{e2.15}
\end{equation}
We choose this enhancement factor since Lacks and Gordon \cite{lacks94} found 
it gave the best approximation to the non-additive kinetic energy of a number 
of related functionals.
The second enhancement factor is
\begin{equation}
F_{TF-\lambda vW}(s)=1+ \lambda \frac{5}{3} s^2,
\label{e2.16}
\end{equation}
which for $\lambda=0$ is the Thomas-Fermi (TF) approximation and for 
$\lambda=\frac{1}{9}$ the gradient expansion truncated at first order.
The parameter $\lambda$ is taken as a free parameter in order to optimise the 
results of the calculations (this is not a completely empirical approach and 
some theoretical justification is available \cite{dreizler90,yang86}).

A number of more general non-local approximations to the kinetic energy 
functional are available, such as local density scaling \cite{ludena98} and 
weighted density approximations (eg Chac\'{o}n et al 
\cite{chacon85},Garc\'{i}a-Gonz\'{a}lez et al \cite{garcia96}).
Although these approaches are accurate they are computationally expensive, 
hence we choose a simpler form which has been found to accurately reproduce 
energies and densities for some bulk systems \cite{wang98}.
This functional is from the family of approximations introduced by Wang and 
Teter \cite{wang92}, Perrot\cite{perrot94} and by Smargiassi and Madden 
\cite{smargiassi94}, summarised and generalised by Wang et al \cite{wang98}.
These take the form
\begin{eqnarray}
T^{nloc}_{\alpha}[\rho]=&&
 \frac{3}{5}(3\pi^2)^{\frac{2}{3}} \int \rho^{\frac{5}{3}} d^3 {\mathbf{r}}
-          \int \rho^{\frac{1}{2}} \nabla^2  \rho^{\frac{1}{2}} d^3 
{\mathbf{r}} \nonumber\\
&& \mbox{} +       \int \rho^{\alpha} w_{\alpha}( {\mathbf{r}} - {\mathbf{r}}') 
\rho^{\alpha} d^3 {\mathbf{r}} d^3 {\mathbf{r}}'
\label{e2.17}
\end{eqnarray}
where the first term is the TF functional, the second term is the von 
Weizsacker (vW) functional and the third term is defined such that the entire 
functional has the correct linear response for a homogenous, non-interacting 
electron gas.
A number of different values have been proposed for the parameter $\alpha$, 
each with its own justification as discussed by Wang et al \cite{wang98}.
For a plane-wave representation of $\rho({\mathbf{r}})$ Eq. (\ref{e2.17}) takes 
the form
\begin{eqnarray}
T^{nloc}_{\alpha}[\rho]=\Omega \sum_{\mathbf{g}} &&
\frac{3}{5}  (3 \pi^2)^{\frac{2}{3}}
                \rho^{\frac{5}{6}}_{\mathbf{g}}     
\rho^{\frac{5}{6}}_{-\mathbf{g}} +
                \rho^{\frac{1}{2}}_{\mathbf{g}} g^2 
\rho^{\frac{1}{2}}_{-\mathbf{g}} \nonumber\\
&&  \mbox{} +          \rho^{\alpha}_{\mathbf{g}}  w_{\alpha}({\mathbf{g}})    
\rho^{\alpha}_{-\mathbf{g}}
\label{e2.18}
\end{eqnarray}
where the powers of the electron density are taken before transformation to 
reciprocal space, and $\Omega$ is the volume of the unit cell.
The linear response correction, $w_{\alpha}(g)$ is given by
\begin{eqnarray}
w_{\alpha}({\mathbf{g}})=&&
           -\frac{k_f^2}{3 \alpha^2 \rho_0^{ 2 \alpha-1 } }
           -\frac{1}    {4 \alpha^2 \rho_0^{ 2 \alpha-1 } } g^2 \nonumber\\
&& \mbox{} -\frac{1}    {2 \alpha^2 \rho_0^{ 2(\alpha-1)} } 
\frac{1}{\chi_{Lind}(g)}
\label{e2.19}
\end{eqnarray}
with $\chi_{Lind}$ the Lindhard susceptibility function in reciprocal space,
\begin{equation}
-\frac{1}{\chi_{Lind}}=\frac{2\pi^2}{k_f} \left (
\frac{1}{2} + \frac{1-\eta^2}{4\eta} \ln 
\left| \frac{1+\eta^2}{1-\eta^2} \right|
\right)^{-1}
\label{e2.20}
\end{equation}
where $k_f=(3\pi^2\rho_0)^\frac{1}{3}$ is the Fermi vector for the average 
electron density $\rho_0$, and $\eta=\frac{g}{2k_f}$.
The functional derivative can be obtained by writing Eq. (\ref{e2.17}) as a 
discrete sum (double sum for the linear response correction) over the real 
space grid, in the same manner as White and Bird's treatment of the semi-local 
functionals.
Introducing an infinitesimal change in $\rho({\mathbf{r}})$ at a grid point 
results in
\begin{eqnarray}
\frac{\delta T^{nloc}_{\alpha}[\rho]}{\delta \rho}= \Omega && \left[
            (3 \pi^2)^{\frac{2}{3}} \rho^{\frac{2}{3}}({\mathbf{r}})
+            \rho^{-\frac{1}{2}}({\mathbf{r}})
\sum_{\mathbf{g}} g^2             \rho^{\frac{1}{2}}_{\mathbf{g}}          
e^{i{\mathbf{g}}.{\mathbf{r}}}  \right. \nonumber\\
&& \left. \mbox{} + 2 \alpha \rho^{\alpha-1}({\mathbf{r}})
\sum_{\mathbf{g}} w_{\alpha}({\mathbf{g}}) \rho^{\alpha}_{\mathbf{g}}           
    e^{i{\mathbf{g}}.{\mathbf{r}}}
\right].
\label{e2.21}
\end{eqnarray}
Equation (\ref{e2.21}) was found to be stable for $\alpha=\frac{1}{2}$, but 
became unstable for other values.
This behaviour can be ascribed to the fact that in the limit of $g \rightarrow 
0$ Eq. (\ref{e2.17}) gives the exact second order gradient expansion only for 
$\alpha=\frac{1}{2}$ \cite{wang98}.
For all other values this not the case, hence the convergence problems 
associated with exponentially decaying, low electron density regions described 
earlier become significant.

\section{Embedding tested on bulk fcc Aluminium}
\label{sec:results}
\subsection{All terms of $T_s^{nadd}$ approximate}
\label{sec:approxrho1}
To investigate this partially frozen density approach we examine fcc aluminium 
with a 4 atom cubic unit cell.
The 3 face-centred atoms are taken to be the substrate system, and this 
structure is solved to provide $\rho_2({\mathbf r})$ and $T_s[\rho_2]$.
A plane-wave basis pseudopotential approach is used, with a lattice constant of 
$a_0=4.05$\AA, a plane-wave cut-off of $200$ eV, $35 {\mathbf k}$ points in the 
irreducible wedge of the Brillouin zone, and the Goodwin-Needs-Heine 
\cite{goodwin90} local pseudopotential (a non-local pseudopotential is 
considered in subsection \ref{sec:nonloc}).
Exchange-correlation is described by the LDA.

Once this substrate is constructed the embedded Kohn-Sham calculation is 
carried out as a standard plane-wave basis calculation, but with the trial 
potential given by Eq. (\ref{e2.5}) and the total energy given by Eq. 
(\ref{e2.2}).
The non-additive kinetic energy in Eq.(\ref{e2.2}) is given by Eq. (\ref{e2.3}) 
with all of the terms on the RHS approximate.
The unit cell, Brillouin zone sampling and other parameters of the calculation 
are chosen to be the same as the substrate calculation.
It should be made clear that for the substrate calculation we are solving for 
the lattice of 3 face centred atoms and their accompanying electrons, but for 
the embedded calculation we are solving for the \emph{entire} fcc system, but 
only the electrons associated with the embedded (corner) atom are provided with 
a Kohn-Sham representation.

\begin{table}
\begin{tabular}{lrrrr} \hline \hline
                        &        &              &          & Peak error/ \\
Functional              & $E$/eV & $\Delta E$/eV&   R/\%   & $\times 10^{-3}$
\AA$^{-3}$ \\ \hline
$T_{TF-\frac{4}{9}vW}$  & $-$58.722 &  $-$0.392 & 4.206 & 28.750   \\ 
$T_{PW86}$              & $-$58.052 &     0.277 & 6.390 & 31.491   \\
$T_{\frac{1}{2}}^{nloc}$& $-$58.337 &  $-$0.008 & 4.614 & 20.051   \\
Kohn Sham               & $-$58.329 &     $-$   &  $-$  &  $-$     \\\hline
\end{tabular}
\caption{Total energy per atom $E$, and errors in energy $\Delta E$ and 
electron density.
Results obtained with embedding scheme described in subsection 
\ref{sec:approxrho1}.}
\label{tab1}
\end{table}

We discuss results for 3 approximate functionals.
Equation (\ref{e2.16}) was applied with $\lambda=\frac{4}{9}$ (denoted 
$T_{TF-\frac{4}{9}vW}$), since this value was found to provide a useful 
compromise between accuracy of the value of the functional itself and the 
functional derivative.
We also employ Eq. (\ref{e2.14}) ($T_{PW86}$) and the non-local linear response 
corrected functional, Eq. (\ref{e2.17}), with $\alpha=\frac{1}{2}$ (denoted 
$T^{nloc}_{ \frac{1}{2}}$).
In Table\ \ref{tab1} the total energy per atom of the embedded calculations are 
given, together with the result of a full Kohn-Sham calculation carried out 
with the same basis and pseudopotential.
$T^{nloc}_{ \frac{1}{2}}$ gives by far the best result for the total energy, 
although the extreme accuracy of this value is probably spurious since similar 
systems consistently provide errors of order $\sim 0.1$ eV/atom.
The total energy for the other functionals is not as accurate, and results for 
other semi-local functionals that are not given here are similar or worse.

\begin{figure}
\begin{center}
\includegraphics*[scale=0.875]{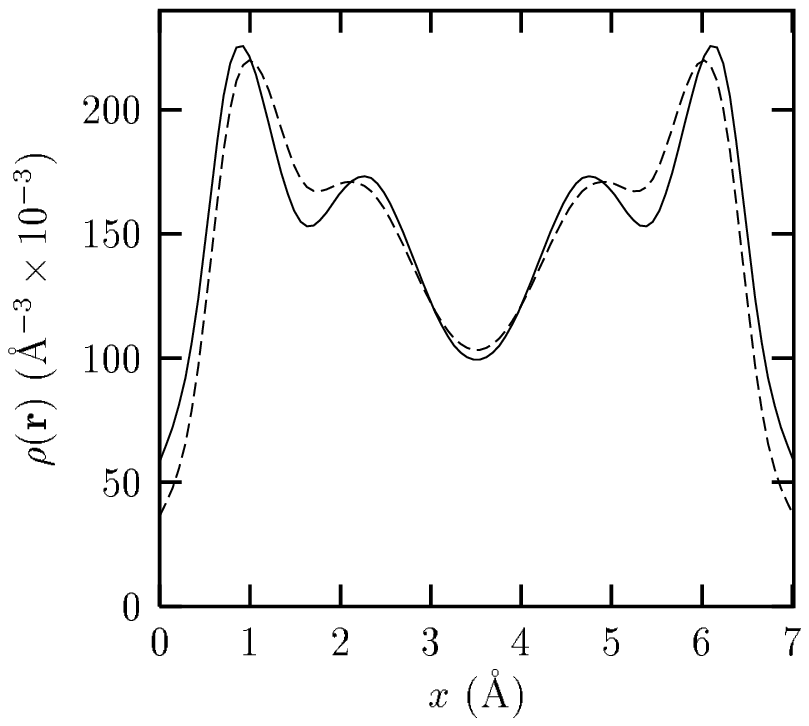}
\includegraphics*[scale=0.875]{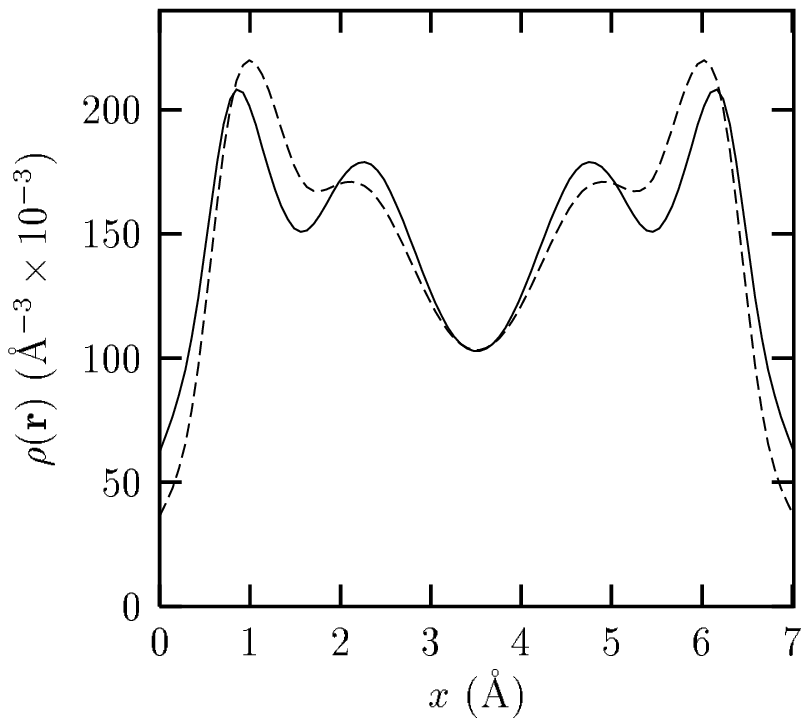}
\includegraphics*[scale=0.875]{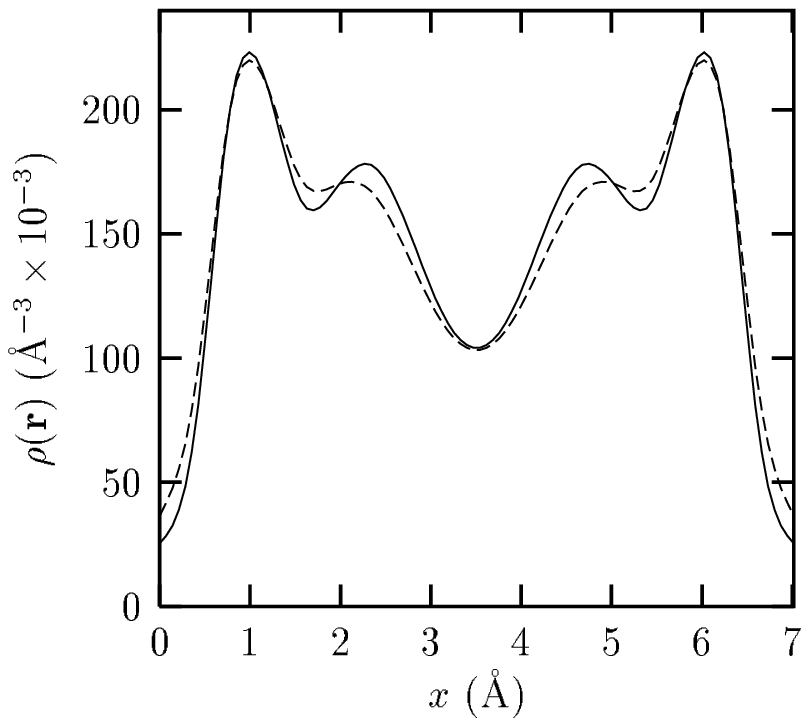} \end{center}
\caption{
Electron density in $[111]$ direction for fcc aluminium using the embedding 
scheme described in subsection \ref{sec:approxrho1}.
Embedded atoms are at $0.00$ and $7.01$ \AA.
The dashed line shows the Kohn-Sham result and the solid line the embedding 
results for a) $T_{TF-\frac{4}{9}vW}$ functional, b) $T_{PW86}$ functional and 
c) $T_{\frac{1}{2}}^{nloc}$ functional (see text).
}
\label{fig1}
\end{figure}

Fig.\ \ref{fig1}a, Fig.\ \ref{fig1}b and Fig.\ \ref{fig1}c show the electron 
density obtained using the $T_{TF-\frac{4}{9}vW}$, $T_{PW86}$ and $T^{nloc}_{ 
\frac{1}{2}}$ functionals respectively.
These figures show the electron density along a line in the $[111]$ direction 
between two embedded atoms on the corners of the cubic unit cell.
Table\ \ref{tab1} gives the error in the electron density, compared with the 
Kohn-Sham results, over the whole unit cell.
This is quantified as the peak error and the average absolute error, $R$, given 
by
\begin{equation}
R=\int |\rho ({\mathbf{r}})-\rho^{KS}({\mathbf{r}})|d^3 {\mathbf{r}} /
  \int                      \rho^{KS} ({\mathbf{r}}) d^3 {\mathbf{r}},
\label{e2.22}
\end{equation}
where $\rho^{KS}({\mathbf{r}})$ is the Kohn-Sham electron density.
From these results it is apparent that the electron densities show the correct 
behaviour, reproducing the structure of the Kohn-Sham electron density 
reasonably well.
However, errors appear near the atomic sites of the embedded atoms as well as 
in in the region where $\rho_1({\mathbf{r}})$ and $\rho_2({\mathbf{r}})$ have 
the greatest overlap.
Considering the errors in both the electron density and total energy, 
$T^{nloc}_{\frac{1}{2}}$ provides the most accurate reproduction of the 
Kohn-Sham results.

\subsection{One term of $T_s^{nadd}$ approximate}
\label{sec:exactrho1}
In this section we approximate the non-additive kinetic energy in a form that 
represents the kinetic energy of the entire system as an approximate kinetic 
energy functional.
However, the densities $\rho_1({\mathbf r})$ and $\rho_2({\mathbf r})$ are 
still represented as Kohn-Sham systems.
This corresponds to the many applications of approximate kinetic energy 
functionals to the direct minimisation of the total energy function (see 
Goedecker \cite{goedecker99}) with no Kohn-Sham representation.

At first representing the electron density as the sum of two Kohn-Sham 
representations may seem like a waste of computational effort since the 
evaluation of the approximate kinetic energy functional for the entire system 
does not require anything more than the electron density itself.
However, it could be useful in two different ways.

First it allows non-local pseudopotentials to be applied directly to each part 
of the system within the Kohn-Sham framework (see subsection 
\ref{sec:nonloc}).
Second the results will tell us whether the error in 
$T_s^{nadd}[\rho_1,\rho_2]$ is greater or smaller than the error in 
$T_s^{app}[\rho_1+\rho_2]$.
This second point is important since for the method described in the previous 
section to be useful the error in the non-additive kinetic energy must be 
smaller than the error in the total kinetic energy, as described by the 
approximate functionals.
Some cancellation of errors must take place in $T_s^{nadd}[\rho_1,\rho_2]$, and 
its functional derivative, for this to be the case.
From the previous subsection it is apparent that cancellation occurs only to a 
limited degree, and a similar conclusion has been reached when addressing the 
accuracy of kinetic energy functionals when used to evaluate interaction 
energies, both for gradient expansions \cite{perdew88} and semi-local 
enhancement factor approximations \cite{lacks94}.

In order to obtain the required functional the non-additive kinetic energy is 
defined as
\begin{equation}
T^{nadd}_s[\rho_1,\rho_2]=
T^{app }_s[\rho_1+\rho_2]-T_s[\rho_1]-T_s[\rho_2],
\label{e3.1}
\end{equation}
where the first term on the RHS is an approximate functional, and the remaining 
terms are exact.
In the previous subsection all of the functionals in this expression were 
evaluated using approximate functionals.
Equation (\ref{e2.2}) is applied as before, with $T_s[\rho_1]$ and 
$T_s[\rho_2]$ exact, but with $T^{nadd}$ given by Eq. (\ref{e3.1}).
The functional derivative of the non-additive kinetic energy becomes
\begin{equation}
\frac{\delta T_s^{nadd}[\rho_1,\rho_2]}{\delta \rho_1} =
\frac{\delta T_s^{app }[\rho_1+\rho_2]}{\delta \rho_1} - \frac{\delta 
T_s[\rho_1]}{\delta \rho_1}.
\label{e3.2}
\end{equation}
where the first term on the RHS is approximate, and the second exact.
To obtain the second term in Eq. (\ref{e3.2}) the method of Bartolotti and 
Acharya \cite{bartolotti82} is applied.
They derive an expression for the functional derivative by replacing the 
self-consistent potential within the Kohn-Sham equations with the self 
consistent potential in terms of the associated Euler-Lagrange equation,
\begin{equation}
- \nabla^2 \psi_n({\mathbf{k}}) + 
\left( \mu'- \frac{\delta T_s[\rho_1]}{\delta \rho_1} \right) 
\psi_n({\mathbf{k}}) =
\epsilon_n({\mathbf{k}}) \psi_n({\mathbf{k}})
\label{e3.3}
\end{equation}
which gives
\begin{equation}
\frac{\delta T_s[\rho_1]}{\delta \rho_1}=
- \frac{ \nabla^2 \psi_n({\mathbf{k}}) }{\psi_n({\mathbf{k}})} 
-\epsilon_n({\mathbf{k}}) +\mu'
\label{e3.4}
\end{equation}
where $\mu'$ is the associated Fermi energy.
It should be noted that the Euler-Lagrange equation can be used directly to 
obtain the functional derivative in terms of the trial potential, but this was 
found to cause convergence difficulties.
To determine $\mu'$ we use
\begin{equation}
T_s[\rho_1]=\int \rho_1({\mathbf{r}}) \frac{\delta T_s[\rho_1]}{\delta \rho_1} 
d^3 {\mathbf{r}},
\label{e3.5}
\end{equation}
as derived by Liu and Parr \cite{liu97}.

\begin{figure}
\begin{center}
\includegraphics*[scale=0.875]{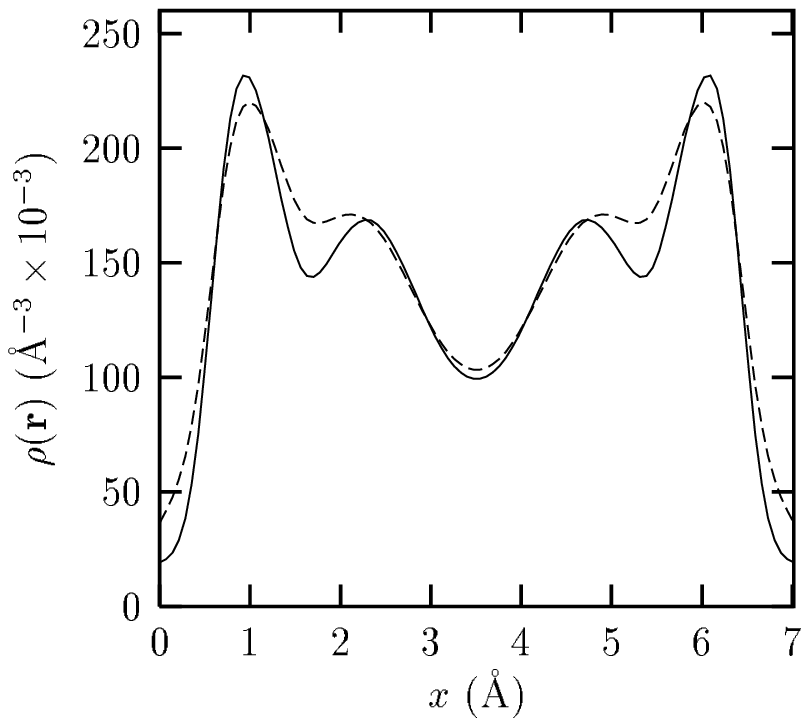}
\includegraphics*[scale=0.875]{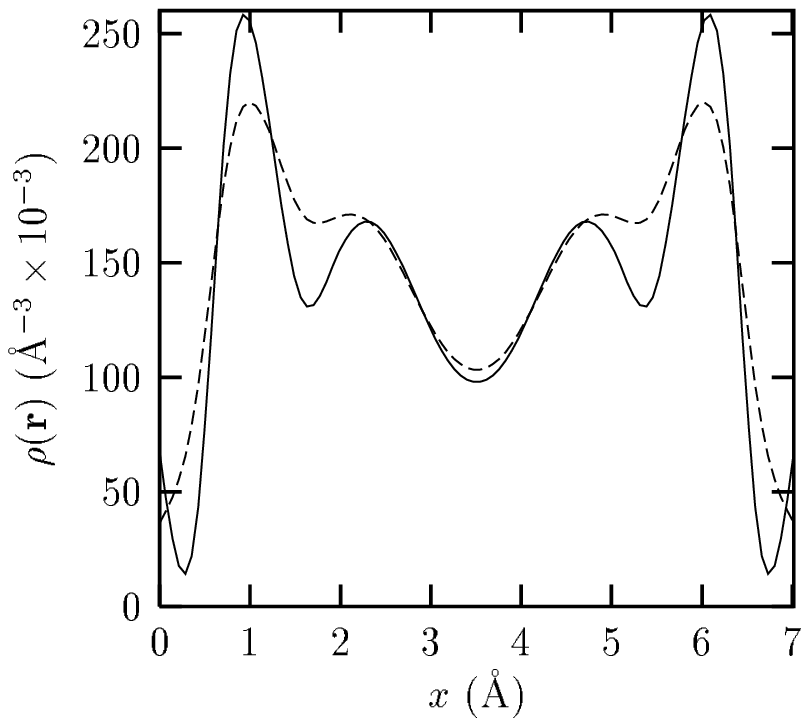}
\includegraphics*[scale=0.875]{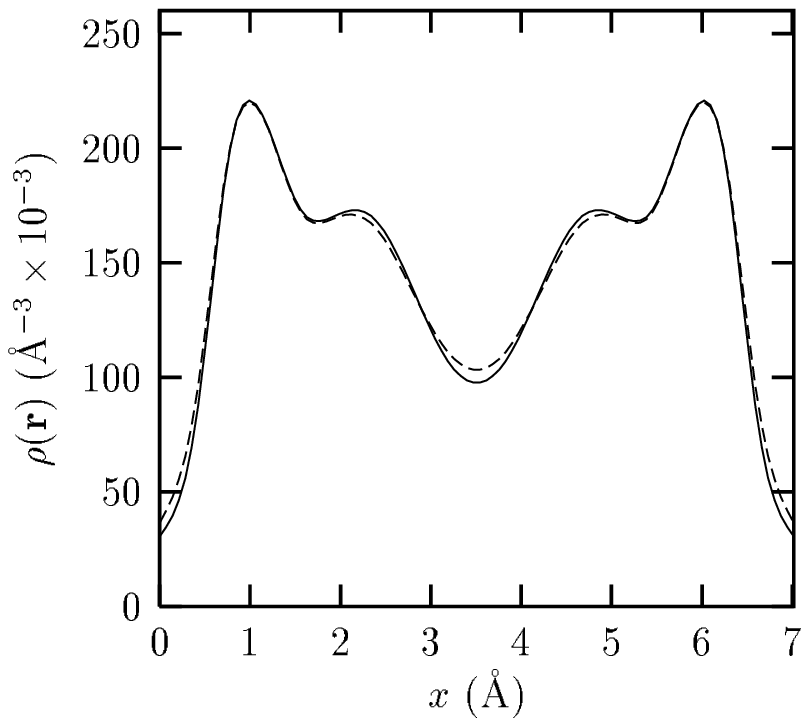} \end{center}
\caption{
Electron density in $[111]$ direction for fcc aluminium using the embedding 
scheme described in subsection \ref{sec:exactrho1}.
Embedded atoms are at $0.00$ and $7.01$ \AA.
The dashed line shows the Kohn-Sham result and the solid line the embedding 
results for a) $T_{TF-\frac{4}{9}vW}$ functional, b) $T_{PW86}$ functional and 
c) $T_{\frac{1}{2}}^{nloc}$ functional (see text).
}
\label{fig2}
\end{figure}

In effect this second approach corresponds to minimising the energy functional 
of the entire system with the kinetic energy represented by the approximate 
kinetic energy functional, but with the additional constraint of $\rho({\mathbf 
r}) \geq \rho_2({\mathbf r})$.
Calculations were carried out within this scheme, with all parameters 
equivalent to those in the previous subsection.
Table\ \ref{tab2} shows the total energy per atom, and the associated errors.
Generally, the errors in the energy for the semi-local functionals are 
considerably worse than in Table\ \ref{tab1}, as demonstrated by the results 
for $T_{PW86}$, but the error for $T_{TF-\frac{4}{9}vW}$ is similar to the 
first method.
The non-local functional gives by far the most accurate results, with the 
energy accurate to better than $0.1$ eV.
Fig.\ \ref{fig2}a, Fig.\ \ref{fig2}b and Fig.\ \ref{fig2}c show the electron 
density resulting from these calculations, for  $T_{TF-\frac{4}{9}vW}$, 
$T_{PW86}$ and $T^{nloc}_{\frac{1}{2}}$ respectively.
It is immediately apparent that the non-local functional provides a far more 
accurate electron density than either of the semi-local functionals, and a more 
accurate electron density than the approach described in subsection 
\ref{sec:approxrho1}.

\begin{table}
\begin{tabular}{lrrrr} \hline \hline
                        &        &              &          & Peak error/ \\
Functional              & $E$/eV & $\Delta E$/eV&   R/\%   & $\times 10^{-3}$
\AA$^{-3}$ \\ \hline
$T_{TF-\frac{4}{9}vW}$  & $-$58.769 & $-$0.440 & 3.922 & 25.050   \\
$T_{PW86}$              & $-$59.680 & $-$1.351 & 7.260 & 57.064   \\
$T_{\frac{1}{2}}^{nloc}$& $-$58.411 & $-$0.082 & 1.204 &  9.180   \\
Kohn Sham               & $-$58.329 &     $-$  &  $-$  &  $-$     \\ \hline
\end{tabular}
\caption{Total energy per atom $E$, and errors in energy $\Delta E$ and  
electron density.
Results obtained with embedding scheme described in subsection 
\ref{sec:exactrho1}.}
\label{tab2}
\end{table}

\begin{figure}
\begin{center} \includegraphics*[scale=0.875]{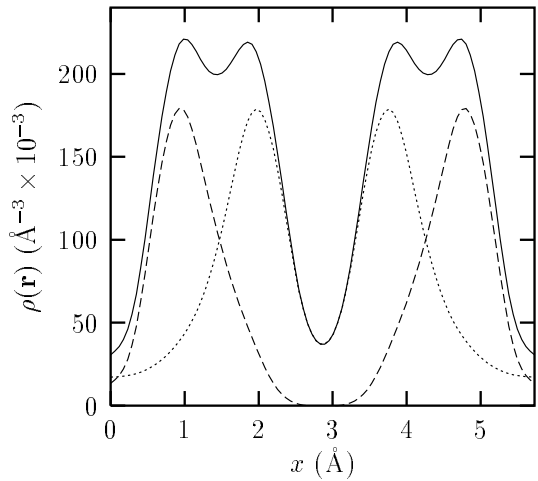} \end{center}
\caption{
Electron density in $[110]$ direction for fcc aluminium using the embedding 
scheme described in subsection \ref{sec:exactrho1} and the 
$T_{\frac{1}{2}}^{nloc}$ functional (see text).
Embedded atoms are at $0.00$ and $5.73$ \AA, and substrate atom is at $2.86$ 
\AA.
The dashed line shows the embedded electron density, $\rho_1$, the dotted line 
the substrate electron density, $\rho_2$, and the solid line the total electron 
density, $\rho=\rho_1+\rho_2$.
}
\label{fig3}
\end{figure}

In subsection \ref{sec:approxrho1} errors in the total energy functional are 
introduced by the \emph{non-additive} part of the kinetic energy, whereas in 
this section the errors are due to the approximate \emph{total} kinetic energy.
Bearing this in mind our results indicate that, for the non-local functional, 
$T[\rho_1+\rho_2]$ is particularly well described by $T^{nloc}_{\frac{1}{2}}$, 
but $T[\rho_1]$ and $T[\rho_2]$ are not quite as accurate (these terms only 
appear in the non-additive kinetic energy).
This difference in the accuracy of the same functional applied to different 
electron densities can be ascribed to the different characters of the electron 
densities.
Fig.\ \ref{fig3} shows embedded, substrate and total electron densities 
resulting from a calculation of the type described in this section, carried 
out with the non-local functional.
Electron densities are shown along a line in the $[110]$ direction between two 
embedded atoms on opposite corners of one face of the cubic unit cell (this 
line is chosen as it includes a substrate atom along its path).
It is apparent that $\rho_1$ is far from homogeneous, and falls close to zero 
near the sites of substrate atoms.
This is expected to be accompanied by worse performance of 
$T^{nloc}_{\frac{1}{2}}[\rho_1]$ than $T^{nloc}_{\frac{1}{2}}[\rho_1+\rho_2]$, 
since this approximate functional is derived using arguments based on the 
linear response of a homogenous electron gas.
It has previously been found that this functional is particularly successful 
for the total electron density of bulk aluminium with the pseudopotential 
applied here \cite{wang98}.
Similar considerations lead us to conclude that in general for semi-local 
functionals the difference 
$T_s^{app}[\rho_1+\rho_2]-T_s^{app}[\rho_1]-T_s^{app}[\rho_2]$ is more accurate 
than $T_s^{app}[\rho_1+\rho_2]$ (or about the same accuracy for the particular 
form $T_{TF-\frac{4}{9}vW}$).

It is worth noting that the results obtained here for Aluminium show closer 
agreement with Kohn-Sham results than those obtained by Wang et al 
\cite{wang98} using a slightly generalised form of the non-local functional, 
the same pseudopotential and direct minimisation of the total energy 
functional.
This is probably due to the additional constraint of $\rho({\mathbf r}) \geq 
\rho_2({\mathbf r})$ present in the calculation performed here preventing an 
over-relaxation of the electron density.
This constraint can be relaxed by swapping the substrate and embedded systems 
after self-consistency has been reached, reaching self consistency again and 
repeating this until convergence of the total system is reached.
Performing this `Freeze and Thaw' \cite{wesolowski93,wesolowski99} procedure 
offers no particular advantage for the test system investigated in this paper 
and when implemented did not influence any of the conclusions given.
The only significant consequence of performing a `Freeze and Thaw' calculation 
was a `drift' of the electron densities away from their associated atoms.
For the non-local and $T_{TF-\frac{4}{9}vW}$ functionals the two electron 
densities remain localised on their host atoms, whereas for the $T_{PW86}$ (and 
other enhancement factor functionals) the two electron densities evolved to 
fill the entire unit cell with no association with the host atoms of 
region $I$ or $II$ apparent.

\subsection{Non-local pseudopotentials}
\label{sec:nonloc}
In past applications of the Hohenberg-Kohn theorem directly to the minimisation 
of the total energy functional the main problem addressed has been the 
inaccuracy of the available approximate kinetic energy functionals.
An additional problem is how to make use of non-local pseudopotentials when no 
Kohn-Sham representation of the electron density is available.
This problem arises due to the Hohenberg-Kohn theorem being strictly applicable 
only for a local external potential \cite{gilbert75,zumbach85}.

For the embedding method applied here no Kohn-Sham representation of the 
\emph{total} electron density is available, so it is not immediately apparent 
how a non-local pseudopotential can be applied.
This issue has been addressed previously, for example Shah et al \cite{shah94} 
provide an \emph{ad hoc} prescription, by taking the square root of the density 
as the basic variable, while Watson et al \cite{watson98} and Anata and Madden 
\cite{anata99} describe a procedure to obtain a new local pseudopotential from 
a non-local one.
Here we make two assumptions.

First, that the Kohn-Sham density matrix $\gamma_s({\mathbf r},{\mathbf r}')$, 
given by
\begin{equation}
\gamma_s({\mathbf r},{\mathbf r}')=
\sum_i w_i \psi^*_i ({\mathbf{r}})\psi_i ({\mathbf{r}}'), 
\end{equation}
(where $w_i$ is the occupation number for state $i$) is an accurate enough 
representation of the actual many-body $1^{st}$ order 
reduced density matrix (this is the assumption implicit in the normal 
application of non-local pseudopotentials within the standard Kohn-Sham scheme).
Assuming this to be a valid approximation the external potential energy is 
given by
\begin{equation}
E_{ext}[\gamma_s({\mathbf{r}},{\mathbf{r}}')]=
\int v({\mathbf{r}}',{\mathbf{r}}) \gamma_s({\mathbf{r}},{\mathbf{r}}') 
d^3{\mathbf{r}}' d^3{\mathbf{r}}
\label{e5.1}
\end{equation}
where $v({\mathbf{r}}',{\mathbf{r}})$ is an external non-local potential.
Second we take advantage of the fact that the electron density component 
$\rho_2({\mathbf{r}})$ is mostly localised near the atomic sites in the 
substrate, and $\rho_1({\mathbf{r}})$ is expected to remain localised around 
the site of the embedded atoms.
Bearing this in mind we take the exact expression
\begin{equation}
\rho({\mathbf{r}})=\rho_1({\mathbf{r}})+\rho_2({\mathbf{r}})
\label{e5.2}
\end{equation}
and generalise it to the approximation
\begin{equation}
\gamma_s  ({\mathbf{r}},{\mathbf{r}}') \approx
\gamma_{s,1}({\mathbf{r}},{\mathbf{r}}') + 
\gamma_{s,2}({\mathbf{r}},{\mathbf{r}}'),
\label{e5.3}
\end{equation}
where $\gamma_{s,1}$ and $\gamma_{s,2}$ are the $1^{st}$ order reduced 
Kohn-Sham density matrices corresponding to electron densities 
$\rho_1({\mathbf{r}})$ and $\rho_2({\mathbf{r}})$.
Of course this can only be exact, over all space, if there is no overlap 
between the electron densities, as can easily be deduced from the fact that 
there is a non-additive component to the kinetic energy functional in the 
first place.

Approximation (\ref{e5.3}) immediately leads to
\begin{equation}
E_{ext}[\gamma_s] \approx
 \int V_{loc}({\mathbf{r}}) \rho({\mathbf{r}}) d^3{\mathbf{r}}
+V_{nloc}[\gamma_{s,1}]+V_{nloc}[\gamma_{s,2}]
\label{e5.4}
\end{equation}
where the first term is the local part of the potential, and the second and 
third terms are the contributions of the embedded and substrate systems due to 
the non-local part of the pseudopotentials of \emph{all} the atoms.
Conventional norm-conserving pseudopotentials are non-local only between points 
on the surface of spheres centred on each atomic site, and only for spheres 
with a radius less than a certain value, $r_c$.
This implies that the approximation in Eq. (\ref{e5.3}) need only be accurate 
between points on each such sphere.
Since $\rho_2({\mathbf{r}})$ is largely localised near the atomic sites in the 
substrate, and $\rho_1({\mathbf{r}})$ is expected to remain localised near the 
embedded atomic sites, Eq. (\ref{e5.4}) provides a reasonable approximation.

\begin{figure}
\begin{center}
\includegraphics*[scale=0.875]{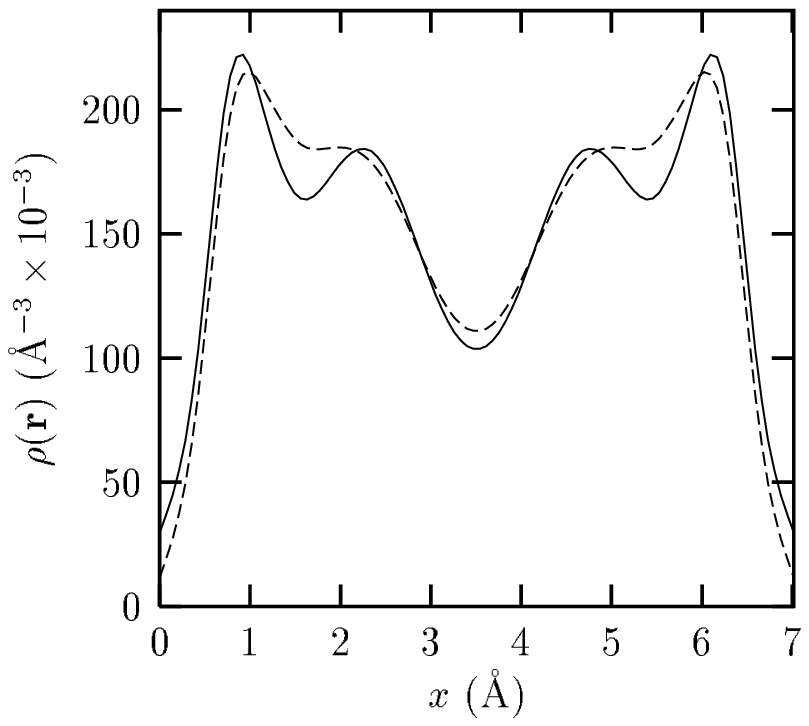}
\includegraphics*[scale=0.875]{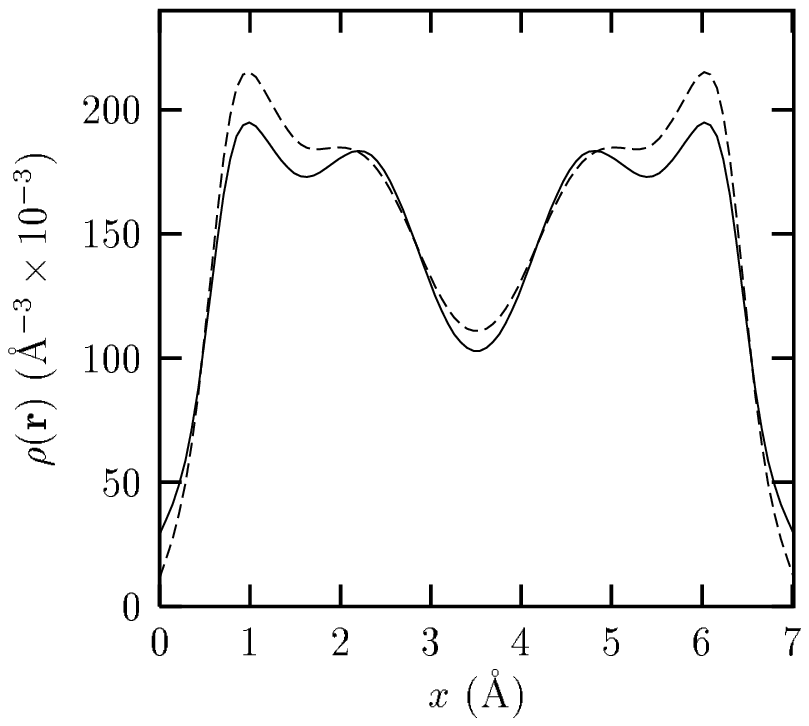}
\includegraphics*[scale=0.875]{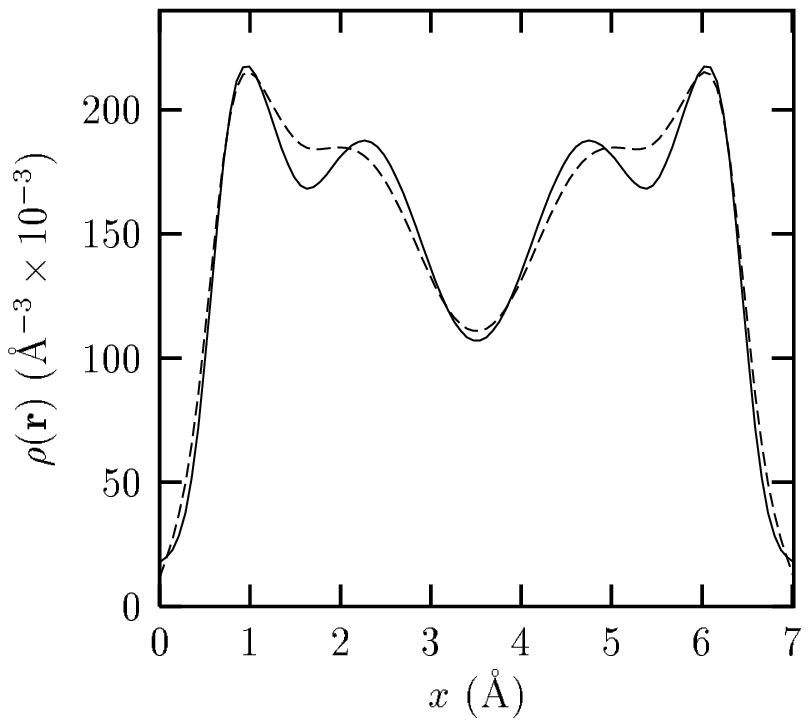} \end{center}
\caption{
Electron density in $[111]$ direction for fcc aluminium using the embedding 
scheme described in subsection \ref{sec:approxrho1} and a non-local 
pseudopotential.
Embedded atoms are at $0.00$ and $7.01$ \AA.
The dashed line shows the Kohn-Sham result and the solid line the embedding 
results for a) $T_{TF-\frac{4}{9}vW}$ functional, b) $T_{PW86}$ functional and 
c) $T_{\frac{1}{2}}^{nloc}$ functional (see text).
}
\label{fig4}
\end{figure}

Calculations were carried out for fcc Al with all parameters as before, and 
using the approach given in subsection \ref{sec:approxrho1} but with a Kerker 
non-local pseudopotential \cite{kerker80}.
Fig.\ \ref{fig4} and Table\ \ref{tab3} show results for a calculation with 
$\rho_2({\mathbf r})$ frozen and for the $T_{TF-\frac{4}{9}vW}$, $T_{PW86}$ and 
non-local functionals.
These results should be compared with those of Fig.\ \ref{fig1} and Table\ 
\ref{tab1}.
For the total energies both $T_{TF-\frac{4}{9}vW}$ and $T^{nloc}_{\frac{1}{2}}$ 
yield results essentially as accurate as those obtained for the local 
pseudopotential in subsection \ref{sec:approxrho1}, while $T_{PW86}$ is less 
accurate.
Fig.\ \ref{fig4} shows a higher peak error in the electron densities, but it 
is difficult to attribute this to a particular aspect of the approximations 
inherent in the implementation of the non-local pseudopotential.
We conclude that a non-local pseudopotential can be used within this method 
with no significant loss of accuracy.

\begin{table}
\begin{tabular}{lrrrr} \hline \hline
                        &        &              &          & Peak error/ \\
Functional              & $E$/eV & $\Delta E$/eV&   R/\%   & $\times 10^{-3}$
\AA$^{-3}$ \\ \hline
$T_{TF-\frac{4}{9}vW}$  & $-$57.148 & $-$0.120 & 4.349 & 25.267   \\
$T_{PW86}$              & $-$56.262 &    0.767 & 7.137 & 41.953   \\
$T_{\frac{1}{2}}^{nloc}$& $-$56.940 &    0.088 & 6.343 & 43.735   \\
Kohn Sham               & $-$57.028 &     $-$   &  $-$  &  $-$    \\ \hline
\end{tabular}
\caption{Total energy per atom $E$, and errors in energy $\Delta E$ and 
electron density.
Results obtained with embedding scheme described in subsection 
\ref{sec:approxrho1}, using a non-local pseudopotential.}
\label{tab3}
\end{table}

\section{Discussion and Conclusion}
\label{sec:conclusion}
We have implemented the partially frozen electron density approach of Cortona 
\cite{cortona91} and Wesolowski and Warshel \cite{wesolowski93,wesolowski94} 
using a plane-wave basis, 
both local and non-local pseudopotentials, and for a metallic system.
Although some numerical instabilities are introduced by using a plane-wave 
representation these are controlled using methods previously developed for 
exchange-correlation energies and potentials.
Several approximations for the kinetic energy functional are considered, 
including the semi-local enhancement factor approximations that have previously 
been applied within this method.
In addition a modified Thomas-Fermi/von Weizsacker functional and a non-local 
functional are implemented.

A lattice of aluminium atoms is embedded into a substrate lattice of atoms to 
create bulk fcc aluminium, and we find that the semi-local functionals result 
in total energies within $\sim 0.2-0.5$ eV per atom of the Kohn-Sham result, 
and the non-local functional results in an energy differing by $\le 0.1$ eV per 
atom.
Kohn-Sham electron densities are reproduced reasonably closely.
The non-local functional performs best with a peak error of $\sim 20$ 
milli-electrons \AA$^{-3}$ while the semi-local functionals result in errors of 
$\sim 30$ milli-electrons \AA$^{-3}$.
Calculations performed with a non-local pseudopotential produce results for the 
total energy and electron density which are of comparable accuracy to the local 
pseudopotential case.

Calculations are also carried out for an altered form of the method where part 
of the kinetic contribution to the embedding potential and energy is obtained 
exactly.
This corresponded to performing a Hohenberg-Kohn minimisation of the total 
energy expressed in terms of the electron density with an approximate kinetic 
energy functional applied to the entire system.
As implemented here this minimisation allows the use of non-local 
pseudopotentials, and introduces the constraint $\rho \geq \rho_2$ where 
$\rho_2({\mathbf{r}})$ is some reference substrate system.
Results are worse than for the true embedding scheme for all the semi-local 
functionals, with the exception of $T_{TF-\frac{4}{9}vW}$ which gives a 
slightly greater error in the energy, but a slightly improved electron density.
From this it seems reasonable to conclude that the approximate non-additive 
kinetic energy is more accurate than the approximate total kinetic energy for 
these functionals.
The non-local functional gives the most accurate results, with the energy 
accurate to $<0.1$ eV atom$^{-1}$ and the electron density accurate to $<10$ 
milli-electrons \AA$^{-3}$.
This suggests that the non-local functional produces the most accurate 
representation of the both the value of the kinetic energy functional and its 
functional derivative, and that the errors in the functional derivative are 
greatest when the electron density is low.

We interpret the extremely good agreement between the Kohn-Sham electron 
density and embedding results found in subsection \ref{sec:exactrho1} 
(see Fig.\ 
\ref{fig2}c) in comparison with that found in subsection \ref{sec:approxrho1} 
(Fig.\ \ref{fig1}c) for the non-local functional to be due to the success of 
this functional in describing bulk aluminium as discussed by Wang et al 
\cite{wang98}.
For systems where the \emph{total} electron density is far from homogeneous (eg 
surface/adsorbate) this success of the method of subsection \ref{sec:exactrho1} 
is not expected to hold.
In future applications the non-local functional and the method of subsection 
\ref{sec:approxrho1} are expected to provide the most accurate reconstruction 
of the full Kohn-Sham result.

\begin{acknowledgements}
This work has been supported by United Kingdom Engineering and Physical 
Sciences Research Council.
We thank T. A. Wesolowski for helpful discussions.
\end{acknowledgements}




\end{document}